\documentclass{amsart}
\usepackage{amssymb}
\usepackage{color}
\begin{document}
\def\nn{\nonumber}
\def\bea{\begin{array}}
\def\eea{\end{array}}
\def\beq{\begin{eqnarray}}
\def\eeq{\end{eqnarray}}
\newtheorem{theorem}{Theorem}[section]
\newtheorem{lemma}[theorem]{Lemma}
\newtheorem{remark}[theorem]{Remark}
\newtheorem{definition}[theorem]{Definition}
\newtheorem{corollary}[theorem]{Corollary}
\newtheorem{example}[theorem]{Example}
\makeatletter
\renewcommand{\theequation}{%
  \thesection.\alph{equation}}
\@addtoreset{equation}{section} \makeatletter
\renewcommand{\theequation}{%
  \thesection.\alph{equation}}
\@addtoreset{equation}{section} \makeatother
\title{Analogies between self-duality and stealth matter source}
\author{Mokhtar Hassa\"{\i}ne}
\begin{address}{Instituto de Matem\'atica y F\'{\i}sica, Universidad de
Talca, Casilla 747, Talca, Chile and Centro~de~Estudios~Cient\'{\i}ficos~(CECS),%
~Casilla~1469,~Valdivia,~Chile.}\end{address}
\begin{email}{hassaine-at-inst-mat.utalca.cl}\end{email}
\begin{abstract}
We consider the problem of a self-interacting scalar field
nonminimally coupled to the three-dimensional BTZ metric such that
its energy-momentum tensor evaluated on the BTZ metric vanishes. We
prove that this system is equivalent to a self-dual system composed
by a set of two first-order equations. The self-dual point is
achieved by fixing one of the coupling constant of the potential in
terms of the nonminimal coupling parameter. At the self-dual point
and up to some boundary terms, the matter action evaluated on the
BTZ metric is bounded below and above. These two bounds are
saturated simultaneously yielding to a vanishing action for
configurations satisfying the set of self-dual first-order
equations.
\end{abstract}
\maketitle

\section{Introduction}
The self-duality is an important notion which has strong
implications from a mathematical as well as physical point of view.
In field theories, the notion of self-duality refers to models for
which the set of second-order field equations can be reduced to a
first-order one. This implies in general to select a particular form
of the interactions and, also to relate the coupling constants of
the problem between them. The resulting first-order equations are in
general more simpler to analyze and correspond to the minimization
of some functionals as the energy or the action.

The most fundamental self-dual model is the four-dimensional
Euclidean Yang-Mills theory for which the action is minimized by the
first-order (anti) self-dual equations
$$
F_{\mu\nu}=\pm
\frac{1}{2}\epsilon_{\mu\nu\lambda\rho}F_{\lambda\rho},
$$
and whose solutions are known as the instantons. In Euclidean
signature the Yang-Mills energy-momentum tensor vanishes identically
for fields satisfying the self-duality equations.

Another important example of self-dual theory is given by the
three-dimensional Abelian-Higgs model which has a minimum energy
self-dual solutions which correspond to vortices. In this case, the
Bogomolnyi self-dual point is defined by $\lambda\propto e^2$, where
$\lambda$ is the coupling constant of the symmetry breaking quartic
potential and $e$ the electric charge of the scalar field. The
functional energy evaluated at this point is bounded and the bound
is satured for fields satisfying the first order self-dual
equations, \cite{Bogomolny:1975de} and \cite{deVega:1976mi}.
However, the first order equations are not solvable and no explicit
solutions are known even in the radial symmetric case. In three
dimensions, (non)relativistic scalar fields minimally coupled to a
gauge field whose dynamics is governed by a Chern-Simons Lagrangian
also exhibit a self-dual behavior, \cite{Hong:1990yh},
\cite{Jackiw:1990aw} and \cite{Jackiw:1990tz}. These self-dual
Chern-Simons theories possess vortex solutions and are relevant in
the study of planar phenomena as the quantum Hall effect or the high
temperature superconductivity. In the (non)relativistic cases, the
self-duality is achieved provided the introduction of a particular
form of the potential with a strength potential fixed in terms of
the Chern-Simons coupling constant and the electric charge of the
scalar field. In both situations, at the self-dual point the energy
is bounded below and the bound is satured by the first-order
self-dual configurations. In contrast with the Yang-Mills model, the
energy-momentum tensor of these self-dual scalar field theories,
namely the Abelian-Higgs model and the Chern-Simons theories, does
not vanish for the self-dual configurations; for detailed reviews on
Chern-Simons self-duality see \cite{Jackiw:1995be},
\cite{Dunne:1998qy} and \cite{Dunne:1995ai}. There exists also
self-dual models obtained by considering at the same time the
Maxwell term and the Chern-Simons term, see e.g. \cite{Lee},
\cite{Manton} and \cite{Hassaine}.

In this paper, we shall be concerned by the problem of a
self-interacting scalar field nonminimally coupled to the
three-dimensional BTZ metric \cite{Banados:1992wn} such that its
energy-momentum tensor $T_{\mu\nu}$ evaluated on the BTZ metric
vanishes. In this case, the set of equations is given by the
conditions $T_{\mu\nu}=0$ evaluated on the BTZ metric while the
scalar field satisfies a non-linear Klein-Gordon equation. This
problem has already been solved by a direct integration of the
second-order field equations \cite{ABMZ}. In this reference, it has
been proved that these configurations, the so-called {\it stealth}
configurations, exist only for a zero angular momentum and for a
particular form of the potential. This later is a local function of
the scalar field and is given by the sum of three different
exponents of the scalar field with two arbitrary coupling constants
while the remaining one is fixed in terms of the nonminimal coupling
parameter.

The purpose of this paper is to establish a certain analogy between
this particular problem of stealth matter and those related to
self-duality models. In particular, we shall prove that the stealth
matter configuration is equivalent to a self-dual system given by a
set of two first-order equations. In the analogy with self-dual
model, the Bogomolnyi self-dual point being achieved by fixing one
coupling of the potential in terms of the nonminimal coupling
parameter. Moreover, we will show that, up to some boundary terms,
the matter action evaluated on the BTZ background is bounded below
and above and both bounds are satured simultaneously at the
self-dual point for configurations satisfying this set of self-dual
first-order equations.

\section{Stealth scalar field over the BTZ black hole}
The fundamental tenet of general relativity is the manifestation of
the curvature of spacetime produced by the presence of matter. This
phenomena is encoded through the Einstein equations that relate the
Einstein tensor (with or without a cosmological constant) to the
energy-momentum tensor of the matter,
\begin{eqnarray}
G_{\mu\nu}+\Lambda g_{\mu\nu}=\kappa\,T_{\mu\nu}.
\label{Einsteineqs}
\end{eqnarray}
Since the energy-momentum tensor depends explicitly on the metric,
both sides of the equations must be solved simultaneously. However,
one can ask the following question: for a fixed geometry solving the
vacuum Einstein equations, is it possible to find a matter source
coupled to this spacetime without affecting it? Concretely, this
problem consists of examining a particular solution of the Einstein
equations (\ref{Einsteineqs}) for which both sides of the equations
vanish, i. e.
\begin{eqnarray}
G_{\mu\nu}+\Lambda g_{\mu\nu}=0=\kappa\,T_{\mu\nu}.
\label{StealthEinsteineqs}
\end{eqnarray}
In three dimensions, such gravitationally undetectable solutions
have been obtained in the context of scalar fields nonminimally
coupled to gravity with a negative cosmological constant \cite{S3d}.
The same problem has also been considered in higher dimensions but
with a flat geometry \cite{Ayon-Beato:2005tu}. More recently, it has
been considered the coupling of gravity to a matter field action for
which the Lagrangian density is a power of the massless Klein-Gordon
Lagrangian. In this case, it has been shown the existence of
nontrivial scalar field configurations with vanishing
energy-momentum tensor on any static, spherically symmetric vacuum
solutions of the Einstein equations \cite{Hassaine:2005xg}.

Here, we are concerned with a nontrivial example of stealth matter
which consists of a three-dimensional self-interacting scalar field
nonminimally coupled to the BTZ black hole metric. The corresponding
action is given by
\begin{eqnarray}\label{eq:ac}
S &=&\int{d^3x}\sqrt{-g} \left(\frac1{2\kappa}\left(R+2l^{-2}\right)
-\frac12\nabla_\mu\Phi\nabla^\mu\Phi -\frac12\xi\,R\,\Phi^2
-U(\Phi)\right),\\
 &= &S_{EH}-S_M,\nonumber
\end{eqnarray}
where $\Lambda=-l^{-2}$ is the cosmological constant, $\xi$ is the
nonminimal coupling parameter and $U(\Phi)$ is the self--interaction
potential that reads
\begin{eqnarray}
U(\Phi)=\lambda_1\Phi^2+\lambda_2\Phi^{(1-2\xi)/\xi}+\lambda_3\Phi^{1/(2\xi)},
\label{pot}
\end{eqnarray}
where $\lambda_1,\lambda_2$ and $\lambda_3$ are arbitrary constants.
The action $S_{EH}$ refers to the Einstein-Hilbert action with
cosmological constant while the matter action $S_M$ is given by
\begin{eqnarray}
S_M (g,\Phi)=\int{d^3x}\sqrt{-g} \left(
\frac12\nabla_\mu\Phi\nabla^\mu\Phi +\frac12\xi\,R\,\Phi^2
+U(\Phi)\right). \label{sm}
\end{eqnarray}
The field equations obtained by varying the metric in the action
(\ref{eq:ac}) are
\begin{equation}\label{eq:Ein}
G_{\mu\nu}-l^{-2}g_{\mu\nu}={\kappa}T_{\mu\nu},
\end{equation}
while the variation of the scalar field yields a generalized
Klein-Gordon equation
\begin{equation}\label{eq:scalar}
\Box\Phi=\xi\,R\,\Phi +\frac{\mathrm{d}U(\Phi)}{\mathrm{d}\Phi},
\end{equation}
where the energy--momentum tensor is given by
\begin{equation}\label{eq:T}
T_{\mu\nu}=\nabla_\mu\Phi\nabla_\nu\Phi
-g_{\mu\nu}\left(\frac12\nabla_\alpha\Phi\nabla^\alpha\Phi+U(\Phi)\right)
+\xi\left(g_{\mu\nu}\Box-\nabla_\mu\nabla_\nu
+G_{\mu\nu}\right)\Phi^2 .
\end{equation}
We posit the metric to be the BTZ metric without angular momentum
\footnote{In Ref. \cite{ABMZ}, it has been shown that the existence
of stealth matter field requires the absence of the angular
momentum. },
\begin{eqnarray}
ds^2=-(\frac{r^2}{l^2}-M)dt^2+(\frac{r^2}{l^2}-M)^{-1}dr^2+r^2
d\theta^2, \label{BTZ}
\end{eqnarray}
where $M$ is the mass of the black hole. Since this metric is a
solution of the $(2+1)$ vacuum Einstein equations with a negative
cosmological constant, our set of equations
(\ref{eq:Ein}-\ref{eq:scalar}) reduce of finding a scalar field
$\Phi$ such that the energy-momentum tensor (\ref{eq:T}) evaluated
on the BTZ metric vanishes, i.e. (\ref{StealthEinsteineqs}).

We now posit a set of two first-order equations,
\begin{subequations}\label{sdb}
\begin{eqnarray}
\label{sdbt} &&\partial_t\Phi=-\frac{2K\xi}{1-4\xi}\sqrt{M
F(r)}\sqrt{\frac{\Big(\Phi^{(4\xi-1)/(2\xi)}-h\Big)^2}{K^2l^2
F(r)}-1}\,\,\Phi^{(1-2\xi)/(2\xi)},\\
\label{sdbr}
&&\partial_r\Phi=-\frac{2\xi}{1-4\xi}\Gamma_{tr}^t\Big(\Phi-h\Phi^{(1-2\xi)/(2\xi)}\Big),
\end{eqnarray}
\end{subequations}
where $F(r)$ is the BTZ metric function, $F(r)=r^2/l^2-M$ and $K$
and $h$ are two constants related to the coupling constant
$\lambda_2$ as
$$
\lambda_2=\frac{2\xi^2}{l^2(1-4\xi)^2}(h^2+K^2l^2M).
$$
After a tedious but straightforward computation, one obtains that
for a scalar field satisfying the set of first-order equations
(\ref{sdb}), the energy-momentum tensor evaluated on the BTZ metric
becomes
\begin{eqnarray}
T_{\mu\nu}=g_{\mu\nu}\Big[\Phi^2\Big(\frac{2\xi^2}{l^2(1-4\xi)}+\lambda_1(4\xi-1)+
\frac{\xi}{l^2}(1-12\xi)\Big)\Big],
\end{eqnarray}
where in order to derive this expression, we have used that the
Christoffel symbols associated to the BTZ metric satisfy
\begin{eqnarray*}
&&(\Gamma^t_{tr})^2=\frac{g_{rr}}{l^2}\Big(1+\frac{M}{F(r)}\Big),\qquad
\partial_r
\Gamma^t_{tr}=-\frac{g_{rr}}{l^2}\Big(1+\frac{2M}{F(r)}\Big),\\
&&\Gamma^t_{tr}\Gamma^r_{rr}=-\frac{g_{rr}}{l^2}\Big(1+\frac{M}{F(r)}\Big),\quad
\Gamma^t_{tr}\Gamma^r_{tt}=-\frac{g_{tt}}{l^2}\Big(1+\frac{M}{F(r)}\Big),\quad
\Gamma^t_{tr}\Gamma^r_{\theta\theta}=-\frac{g_{\theta\theta}}{l^2}.
\end{eqnarray*}
Hence, for scalar fields satisfying the set of first-order equations
(\ref{sdb}), the condition $T_{\mu\nu}=0$ is verified provided the
coupling constant $\lambda_1$ to be fixed in terms of the nonminimal
coupling parameter as
\begin{eqnarray}
\lambda_1=\frac{\xi}{l^2(1-4\xi)^2}(1-8\xi)(1-6\xi),\label{sdbp}
\end{eqnarray}
while the third coupling constant $\lambda_3$ is fixed by solving
the generalized Klein-Gordon equation. Note that the relation
(\ref{sdbp}) is analogue to the Bogomolnyi self-dual point where the
coupling constants of the problem are in general related between
them. The resulting potential becomes exactly the one derived in
\cite{ABMZ},
\begin{eqnarray}
U(\Phi)=\frac{\xi}{l^2(1-4\xi)^2}\Big((1-8\xi)(1-6\xi)\Phi^2+
2\xi(h^2+K^2l^2M)\Phi^{(1-2\xi)/\xi}\nonumber\\
+4\xi(1-8\xi)h\Phi^{1/(2\xi)}\Big).\label{pott}
\end{eqnarray}
Various comments can be made concerning the structure of this
potential. Firstly, for the three-dimensional conformal coupling
$\xi=1/8$, this potential reduces to the conformal one in three
dimensions. Secondly, at the vanishing cosmological constant limit
($l\to\infty$), we recover the potential allowing a self-interacting
scalar field to be nonminimally coupled to a \emph{pp} wave
background \cite{Ayon-Beato:2005bm}. It is intriguing that this
potential belongs to the same family of potentials arising in the
context of scalar fields nonminimally coupled to special geometries
without inducing back reaction (the static BTZ black hole
\cite{ABMZ} or in flat space \cite{Ayon-Beato:2005tu}). Recently it
has also been shown that this potential arises in the context of
scalar fields nonminimally coupled to an AdS wave geometry
\cite{Ayon-Beato:2005qq}.

In sum, we have shown that the scalar configuration given by a
scalar field satisfying the self-dual equations (\ref{sdb}) together
with the potential (\ref{pott}) is also solution of the stealth
equations evaluated on the BTZ metric
(\ref{eq:Ein}-\ref{eq:scalar}). It is easy to solve the self-dual
equations, and the solution is given by
\begin{eqnarray}
\Phi(t,r)=\Big(\cosh(\sqrt{M}t/l)\sqrt{r^2-Ml^2}+h\Big)^{2\xi/(4\xi-1)}
\end{eqnarray}
which agrees with the solution obtained in \cite{ABMZ}.

As said in the introduction, the self-dual equations in self-dual
models correspond in general to the minimization of some functionals
of the problem as the energy or the action. In our case, we prove
that the matter action evaluated on the BTZ metric is bounded below
and above up to some surface terms and these two bounds are achieved
in the self-dual point (\ref{sdbp}) for configurations satisfying
the self-dual equations (\ref{sdb}).

The matter action (\ref{sm}) evaluated on the BTZ metric at the
self-dual point (\ref{sdbp}) reads
\begin{eqnarray}
\label{ganso} S_M=\int
r\Big(-\frac{1}{2F(r)}(\partial_t\Phi)^2+\frac{1}{2}F(r)(\partial_r\Phi)^2
-\frac{3\xi}{l^2}\Phi^2+U(\Phi)\Big)dr dt d\theta,
\end{eqnarray}
where $U$ refers to the self-dual potential (\ref{pott}) and the
scalar curvature is given by $R=-6/l^2$. It is interesting to note
that the functional (\ref{ganso}) can be rewritten as
\begin{eqnarray}
\label{gansoo} &&\tilde{S}_M=2\pi \int dr dt\Big\{\frac{1}{2}r
F(r)\Big[\partial_r\Phi+\frac{2\xi}{1-4\xi}\Gamma_{tr}^t\Big(\Phi-h\Phi^{(1-2\xi)/(2\xi)}\Big)\Big]^2\nonumber\\
&&-\frac{r}{2 F(r)}\left[\partial_t\Phi+\frac{2K\xi}{1-4\xi}\sqrt{M
F(r)}\sqrt{\frac{(\Phi^{(4\xi-1)/(2\xi)}-h)^2}{K^2l^2
F(r)}-1}\,\,\Phi^{(1-2\xi)/(2\xi)}\right]^2\Big\},\nonumber\\
\end{eqnarray}
where in order to derive this expression we have first operated some
integrations by parts and then dropping all the surface terms. From
the expression (\ref{gansoo}), it is easy to see that the functional
$\tilde{S}_M$ is bounded below and above and both bounds are
simultaneously saturated yielding to a vanishing action, i.e.
$\tilde{S}_M=0$, for configurations satisfying the set of self-dual
equations (\ref{sdb}).

\section{Conclusions}
Here, we have addressed the problem of the analogy that may exist
between self-duality and a stealth matter configuration. We have
considered the problem of a self-interacting scalar field
nonminimally coupled to the BTZ metric. The resulting stealth matter
field equations reduce to the vanishing of the energy-momentum
tensor expressed on the BTZ metric together with a generalized
Klein-Gordon equation. The potential is given by the sum of three
different exponents of the scalar field with two arbitrary coupling
constants while the remaining one is fixed in terms of the
nonminimal coupling parameter. We have exhibited some analogy
between this problem and those arising in the context of self-dual
theories. In particular, the coupling constant of the potential
fixed in terms of the nonminimal coupling parameter is shown to play
the role of the Bogomolnyi self-dual point. In addition, we have
exhibited a set of first-order equations and prove that any scalar
field satisfying these self-dual equations at the self-dual point is
also solution of the stealth equations. Moreover, we have shown that
at the self-dual point, the matter action evaluated on the BTZ
metric is bounded below and above, up to some surface terms that we
have dropped. Curiously enough, these two bounds are satured
simultaneously yielding to a vanishing action for configurations
satisfying the set of self-dual first-order equations.

In the self-dual models, the problem of the existence and uniqueness
of the self-dual solutions is a nontrivial issue. For example, it is
well-known that in the Abelian Higgs model, all the finite energy
solutions of the full second-order field equations are also
solutions to the first-order self-duality equations. In our case,
due to the uniqueness of the solution of the starting problem
\cite{ABMZ}, one can also conclude to the uniqueness of the
self-dual solution.

In order to have a more deep connection between the self-duality and
the stealth configuration, it would be desirable to explore the
issue of the supersymmetric extension. Indeed, in the self-dual
models, the self-dual point is also the point at which the theory
can be extended to a supersymmetric model. An interesting work will
consist to see wether the self-interacting scalar field nonminimally
coupled to the BTZ metric can be extended to a supersymmetric
theory.

\section*{Acknowledgements}
We thank an anonymous referee for calling our attention on this
problem. We always grateful useful discussions with the staff of the
CECS. This work is partially supported by grants 1051084 and 1060831
from FONDECYT. Institutional support to the Centro de Estudios
Cient\'{\i}ficos (CECS) from Empresas CMPC is gratefully
acknowledged. CECS is a Millennium Science Institute and is funded
in part by grants from Fundaci\'{o}n Andes and the Tinker
Foundation.



\begin{thebibliography}{10}

\bibitem{Bogomolny:1975de} E.~B.~Bogomolny, Sov.\ J.\ Nucl.\ Phys.\  {\bf 24}, 449
(1976), [Yad.\ Fiz.\  {\bf 24}, 861 (1976)].

\bibitem{deVega:1976mi} H.~J.~de Vega and F.~A.~Schaposnik, Phys.\ Rev.\ D {\bf 14}, 1100 (1976).


\bibitem{Hong:1990yh} J.~Hong, Y.~Kim and P.~Y.~Pac, Phys.\ Rev.\ Lett.\  {\bf 64}, 2230 (1990).

\bibitem{Jackiw:1990aw} R.~Jackiw and E.~J.~Weinberg, Phys.\ Rev.\ Lett.\  {\bf 64}, 2234 (1990).

\bibitem{Jackiw:1990tz} R.~Jackiw and S.~Y.~Pi, Phys.\ Rev.\ Lett.\  {\bf 64}, 2969 (1990);
Prog.\ Theor.\ Phys.\ Suppl.\  {\bf 107}, 1 (1992).


\bibitem{Jackiw:1995be} R.~Jackiw, {\it Diverse topics in theoretical and mathematical
physics}, World Scientific.

\bibitem{Dunne:1998qy} G.~V.~Dunne, Aspects of Chern-Simons theory, arXiv:hep-th/9902115.

\bibitem{Dunne:1995ai} G.~V.~Dunne, Self-dual Chern-Simons theories,
Lect.\ Notes Phys.\  {\bf M36}, 1 (1995); arXiv:hep-th/9410065.


\bibitem{Lee} C.~K.~Lee, K.~M.~Lee and H.~Min, Phys.\ Lett.\ B {\bf 252}, 79 (1990).

\bibitem{Manton} N.~S.~Manton, Annals Phys.\  {\bf 256}, 114 (1997)

\bibitem{Hassaine} M.~Hassaine, P.~A.~Horvathy and J.~C.~Yera, Annals Phys.\ {\bf 263}, 276 (1998).


\bibitem{Banados:1992wn} M.~Banados, C.~Teitelboim and J.~Zanelli, Phys.\ Rev.\ Lett.\  {\bf 69}, 1849
(1992).



\bibitem{ABMZ}{E. Ay\'on-Beato, C. Mart\'{\i}nez and
J. Zanelli}, Gen. Rel. Grav. {\bf 38}, 145 (2006).



\bibitem{S3d} M.~Natsuume, T.~Okamura, and M.~Sato, Phys.\ Rev.\ D \textbf{61}, 104005
(2000); E.~Ay\'{o}n-Beato, A.~Garc\'{\i}a, A.~Mac\'{\i}as, and
J.M.~P\'{e}rez-S\'{a}nchez, Phys.\ Lett.\ B \textbf{495}, 164
(2000); M.~Henneaux, C.~Mart\'{\i}nez, R.~Troncoso, and J.~Zanelli,
Phys.\ Rev.\ D \textbf{65}, 104007 (2002); J.~Gegenberg,
C.~Mart\'{\i}nez, and R.~Troncoso, Phys.\ Rev.\ D \textbf{67},
084007 (2003).


\bibitem{Ayon-Beato:2005tu} E.~Ayon-Beato, C.~Martinez, R.~Troncoso and J.~Zanelli, Phys.\ Rev.\ D {\bf 71},
104037 (2005).







\bibitem{Hassaine:2005xg} M.~Hassaine, J. Math. Phys. {\bf 47}, 033101 (2006).


\bibitem{Ayon-Beato:2005bm} E.~Ayon-Beato and M.~Hassaine, Phys.\ Rev.\ D {\bf 71}, 084004
(2005).


\bibitem{Ayon-Beato:2005qq} E.~Ayon-Beato and M.~Hassaine, Phys.\ Rev.\ D {\bf 73}, 104001 (2006).









\end{thebibliography}
\end{document}